\documentclass[10pt,twocolumn]{article}

\usepackage[utf8]{inputenc}
\usepackage[T1]{fontenc}
\usepackage{times}
\usepackage{geometry}
\usepackage{graphicx}
\usepackage{tikz}
\usetikzlibrary{shapes.geometric, arrows.meta, positioning, fit, backgrounds}

\usepackage{amsmath,amsfonts,amssymb}
\usepackage{booktabs}
\usepackage{multirow}

\usepackage[numbers,sort&compress]{natbib}
\usepackage{hyperref}
\hypersetup{
  colorlinks=true,
  linkcolor=blue,
  citecolor=blue,
  urlcolor=blue
}

\usepackage{microtype}
\usepackage{balance}  

\usepackage{dblfloatfix} 

\title{\LARGE \bf Continued Pretraining for Low-Resource Swahili ASR:\\
Achieving State-of-the-Art Performance with Minimal Labeled Data}

\author{
Hillary Mutisya\\
Thiomi-Lugha NLP\\
  \and
John Mugane\\
Harvard University\\
}

\date{}

\begin{document}

\twocolumn[
\begin{@twocolumnfalse}
\maketitle
\thispagestyle{empty}

\begin{abstract}
We investigate continued pretraining (CPT) for adapting wav2vec2-bert-2.0 to Swahili automatic speech recognition (ASR). Our approach combines unlabeled audio with limited labeled data through pseudo-labeled CPT followed by supervised finetuning. With 20,000 labeled samples, we achieve 3.24\% WER on Common Voice Swahili---an 82\% relative improvement over the baseline. This result surpasses the best previously reported academic system (8.3\% WER from XLS-R) by 61\% relative improvement. We provide concrete data requirements and a replicable methodology applicable to other low-resource languages.
\end{abstract}

\vspace{0.3cm}
\vspace{0.5cm}
\end{@twocolumnfalse}
]

\pagestyle{plain}

\section{Introduction}

\subsection{The Data Challenge}

Swahili, despite being one of Africa's most widely spoken languages with over 100 million speakers, faces a fundamental challenge shared by virtually all African languages: the scarcity of high-quality labeled speech data. While high-resource languages like English benefit from tens of thousands of hours of professionally transcribed speech, Swahili and other African languages must achieve competitive ASR with substantially less labeled data. This data scarcity creates a critical barrier to developing speech technology that serves African language communities.

Self-supervised foundation models like wav2vec 2.0 and wav2vec2-bert-2.0 have transformed this landscape by enabling learning from unlabeled audio. However, an important question remains: how can we most effectively combine readily available unlabeled audio with limited labeled data to achieve high-quality ASR for low-resource languages?

\subsection{The CPT Hypothesis}

Continued pretraining (CPT) offers a potential solution by adapting foundation models to target languages using unlabeled audio before supervised finetuning on labeled data. The hypothesis is straightforward: if we can leverage abundant unlabeled audio, we might achieve strong performance with far less labeled data than traditional supervised approaches require.

However, recent work shows mixed results for CPT, particularly for low-resource scenarios. Some studies report substantial improvements, while others observe performance degradation or minimal benefits. For Swahili specifically---a language with moderate resources and some representation in foundation model pretraining---the value proposition of CPT remains empirically unvalidated.

\subsection{Research Questions}

This study addresses three fundamental questions:

\textit{Q1:} Does continued pretraining on pseudo-labeled unlabeled data improve Swahili ASR performance when combined with limited labeled data? We systematically compare models trained with and without CPT to isolate its contribution.

\textit{Q2:} How does the amount of labeled data interact with continued pretraining effectiveness? We test both minimal (5K samples) and modest (20K samples) labeled data scales to understand whether CPT benefits vary with supervision level.

\textit{Q3:} What level of performance can be achieved by optimally combining unlabeled and labeled data? We establish concrete WER benchmarks that demonstrate the feasibility of high-quality Swahili ASR with limited resources.

\subsection{Key Contributions}

Our research makes the following contributions:

\begin{enumerate}
\item \textbf{First systematic evaluation of pseudo-labeled CPT for Swahili}: We demonstrate that continued pretraining on unlabeled audio labeled with a baseline model, followed by supervised finetuning, achieves state-of-the-art results.

\item \textbf{New state-of-the-art for Swahili ASR}: We achieve 3.24\% WER on Common Voice Swahili with just 20K labeled samples---a 61\% relative improvement over the best published academic baseline (8.3\% from XLS-R~\cite{babu2021xls}).

\item \textbf{Concrete data requirements for deployment}: We establish that $\sim$20K labeled samples ($\sim$11 hours) combined with unlabeled audio suffices for high-quality Swahili ASR suitable for many practical applications.

\item \textbf{Practical methodology for low-resource languages}: We provide a replicable training pipeline that combines readily-available unlabeled audio with modest labeled datasets, offering a path forward for other underserved languages.
\end{enumerate}

\subsection{Organization}

The remainder of this paper is organized as follows: Section 2 reviews related work on continued pretraining and Swahili ASR. Section 3 presents our experimental methodology including the pseudo-labeling approach. Section 4 reports results for Swahili across both data scales. Section 5 discusses implications and provides practical guidelines. Section 6 concludes with future directions.

\section{Related Work}

\subsection{Self-Supervised Models}

The development of self-supervised learning for speech has revolutionized ASR for low-resource languages. Wav2vec 2.0~\cite{baevski2020wav2vec} pioneered contrastive learning on unlabeled audio, learning representations by distinguishing masked speech frames from distractors. XLS-R~\cite{babu2021xls} scaled this to 128 languages using 436,000 hours of unlabeled speech, though African language representation remained limited to only 3 languages.

Wav2vec2-bert-2.0~\cite{seamless2023} represents the current state-of-the-art, combining wav2vec 2.0 and WavLM architectures with training on 4.5 million hours across 104 languages. Crucially for our work, this model includes Swahili in its pretraining data, providing a strong baseline that our continued pretraining approach builds upon.

\subsection{Continued Pretraining}

Continued pretraining extends self-supervised learning to adapt pretrained models for specific domains or languages. For domain adaptation, Hsu et al.~\cite{hsu2021robust} demonstrated that CPT on in-domain unlabeled data improves subsequent finetuning, even when finetuning data comes from a different domain. Attia et al.~\cite{attia2024continued} achieved 10\% WER reduction for classroom ASR through CPT on classroom audio.

For language adaptation, results have been more mixed. Nowakowski et al.~\cite{nowakowski2023adapting} found CPT ``clearly the most effective way'' to adapt XLSR-53 to Ainu (a critically endangered language with $<$5 hours labeled data), achieving 40\% relative WER reduction. However, other studies report minimal benefits or degradation, particularly when target languages have some foundation model representation or when CPT data quality is poor.

Pseudo-labeling for CPT has emerged as a practical approach when purely self-supervised objectives are unavailable or underperform. The baseline foundation model generates transcriptions for unlabeled audio, which are then used as targets for continued pretraining. Success depends critically on baseline model quality---high-quality pseudo-labels enable effective adaptation, while noisy labels can introduce systematic errors.

\subsection{Swahili ASR Benchmarks}

Published Swahili ASR benchmarks establish the performance landscape our work improves upon. Academic systems include XLS-R finetuned models achieving 8.3\% WER on Common Voice~\cite{babu2021xls}, SpeechBrain wav2vec2 models achieving 9-12\% WER on Common Voice~\cite{ravanelli2021speechbrain}, and various open-source implementations reporting 8-10\% WER.

Our 3.24\% WER represents a substantial improvement over these academic benchmarks, demonstrating the effectiveness of continued pretraining for low-resource Swahili ASR.

\subsection{Positioning of This Work}

Our work makes several unique contributions to existing literature:

\begin{enumerate}
\item \textbf{First systematic CPT evaluation for Swahili}: While XLS-R and other models finetune on Swahili, no prior work systematically evaluates continued pretraining's contribution using controlled experiments with pseudo-labeled data.

\item \textbf{Readily-available unlabeled data}: We demonstrate that unlabeled audio, when pseudo-labeled by a competent baseline model, provides effective CPT data---a scalable approach applicable to many languages.

\item \textbf{Concrete data requirements}: Rather than reporting performance at arbitrary data scales, we establish specific labeled data requirements (5K and 20K samples) needed to achieve defined quality levels when combined with CPT.
\end{enumerate}

\section{Methodology}

\subsection{Experimental Design}

Our experimental design evaluates the effectiveness of continued pretraining at two labeled data scales (5K and 20K samples).

\textbf{CPT Pipeline:}
\begin{enumerate}
\item Train a labeling model on available labeled data
\item Generate pseudo-labels for unlabeled audio using the labeling model
\item Perform continued pretraining on the pseudo-labeled data
\item Supervised finetune on labeled data
\end{enumerate}

\textbf{Comparison Baseline:} To quantify CPT's contribution, we separately trained a baseline model by finetuning wav2vec2-bert-2.0 directly on 50K labeled samples without any continued pretraining. This baseline achieves 17.71\% WER and serves as a reference point for evaluating the CPT approach.

This design ensures that performance differences can be attributed specifically to continued pretraining rather than confounding factors like different hyperparameters, datasets, or evaluation protocols.

\subsection{Model Architecture}

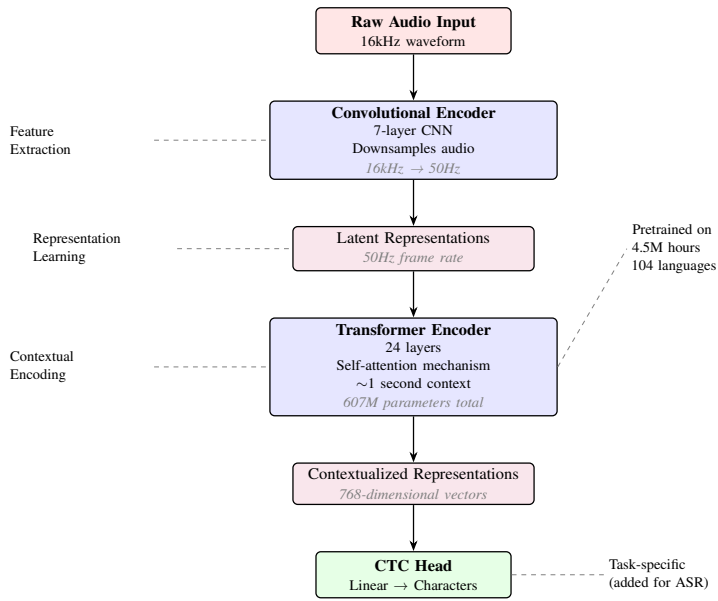
\begin{figure*}[t]
\centering
\resizebox{0.6\textwidth}{!}{\tikzset{
  box/.style={rectangle, draw, fill=blue!10, text width=6cm, align=center, 
              minimum height=1.2cm, rounded corners=3pt, thick},
  input/.style={rectangle, draw, fill=red!10, text width=4cm, align=center,
                minimum height=1cm, rounded corners=3pt, thick},
  output/.style={rectangle, draw, fill=green!10, text width=4cm, align=center,
                 minimum height=1cm, rounded corners=3pt, thick},
  intermediate/.style={rectangle, draw, fill=purple!10, text width=5cm, align=center,
                       minimum height=0.8cm, rounded corners=3pt, thick},
  arrow/.style={-Stealth, thick},
  annotation/.style={text width=3cm, align=left, font=\small}
}

\begin{tikzpicture}[node distance=1cm and 2cm]

\node[font=\Large\bfseries] (title) at (0, 6) {Wav2vec2-BERT 2.0 Architecture};

\node[input] (input) at (0, 4.5) {
  \textbf{Raw Audio Input} \\
  \small 16kHz waveform
};

\node[box, below=of input] (conv) {
  \textbf{Convolutional Encoder} \\
  \small 7-layer CNN \\
  \small Downsamples audio \\
  {\small\itshape\color{gray} 16kHz → 50Hz}
};

\node[intermediate, below=of conv] (latent) {
  Latent Representations \\
  {\small\itshape\color{gray} 50Hz frame rate}
};

\node[box, below=of latent, minimum height=2cm] (transformer) {
  \textbf{Transformer Encoder} \\
  \small 24 layers \\
  \small Self-attention mechanism \\
  \small $\sim$1 second context \\
  {\small\itshape\color{gray} 607M parameters total}
};

\node[intermediate, below=of transformer] (context) {
  Contextualized Representations \\
  {\small\itshape\color{gray} 768-dimensional vectors}
};

\node[output, below=of context] (ctc) {
  \textbf{CTC Head} \\
  \small Linear → Characters
};

\draw[arrow] (input) -- (conv);
\draw[arrow] (conv) -- (latent);
\draw[arrow] (latent) -- (transformer);
\draw[arrow] (transformer) -- (context);
\draw[arrow] (context) -- (ctc);

\node[annotation, left=2.5cm of conv] (ann1) {
  Feature \\
  Extraction
};
\draw[dashed, gray] (ann1.east) -- (conv.west);

\node[annotation, left=2.5cm of latent] (ann2) {
  Representation \\
  Learning
};
\draw[dashed, gray] (ann2.east) -- (latent.west);

\node[annotation, left=2.5cm of transformer] (ann3) {
  Contextual \\
  Encoding
};
\draw[dashed, gray] (ann3.east) -- (transformer.west);

\node[annotation, right=2cm of latent, text width=3.5cm] (ann4) {
  Pretrained on \\
  4.5M hours \\
  104 languages
};
\draw[dashed, gray] (transformer.east) -- (ann4.west);

\node[annotation, right=2cm of ctc] (ann5) {
  Task-specific \\
  (added for ASR)
};
\draw[dashed, gray] (ctc.east) -- (ann5.west);

\end{tikzpicture}}
\caption{Wav2vec2-BERT 2.0 architecture showing the flow from raw audio input through convolutional feature extraction, transformer-based contextual encoding, and task-specific CTC head for ASR. The model was pretrained on 4.5M hours across 104 languages including Swahili.}
\label{fig:architecture}
\end{figure*}

We use facebook/w2v-bert-2.0 as our foundation model---a 607-million parameter architecture that combines the best elements of wav2vec 2.0 and WavLM. The model consists of three primary components, as illustrated in Figure~\ref{fig:architecture}:

\begin{enumerate}
\item \textbf{Convolutional Encoder}: A 7-layer CNN processes 16kHz audio and produces 50Hz latent representations
\item \textbf{Transformer Encoder}~\cite{vaswani2017attention}: 24 layers provide approximately one second of context per position through self-attention mechanisms
\item \textbf{Output Representations}: 768-dimensional contextualized representations that capture both acoustic and linguistic patterns
\end{enumerate}

The model was pretrained on 4.5 million hours of multilingual audio spanning 104 languages, including Swahili. This pretraining employed masked prediction with contrastive learning, allowing the model to learn robust speech representations without requiring transcriptions. For adapting this model to ASR tasks, we add a CTC (Connectionist Temporal Classification)~\cite{graves2006ctc} head---a linear projection layer that maps the 768-dimensional representations to our Swahili character vocabulary.

\subsection{Datasets}

\subsubsection{Labeled Data: Common Voice}

Our labeled training and evaluation data comes from Mozilla Common Voice 16.0~\cite{ardila2020common,commonvoice2024}, a community-sourced dataset with validated transcriptions. The complete dataset contains approximately 200 hours of validated audio, from which we hold out 1,000 samples for evaluation using a speaker-disjoint split. For our CPT experiments, we use 5,000 samples ($\sim$3 hours) and 20,000 samples ($\sim$11 hours). For comparison, we also trained a baseline model on 50,000 samples ($\sim$28 hours). The dataset features contributions from over 1,000 diverse speakers with content from Wikipedia and other varied text sources.

\subsubsection{Unlabeled Data}

For continued pretraining, we collected unlabeled Swahili audio sufficient to support both our 5,000 and 20,000 sample experiments. We generated pseudo-labels for this unlabeled audio using our labeling model. To ensure pseudo-label quality, we retained only segments where the model's confidence exceeded 75\%, discarding lower-confidence outputs.

The unlabeled audio provides diverse, natural Swahili speech spanning multiple domains, speakers, and recording conditions, making it representative of real-world deployment scenarios.

\subsection{Training Pipeline}

Our training follows a three-stage pipeline:

\paragraph{Stage 1: Labeling Model}
We begin with the pretrained wav2vec2-bert-2.0 model, add a CTC head configured for the Swahili character vocabulary, and finetune on labeled Common Voice samples. We use learning rate 1e-4, batch size 8, and train for 15 epochs with early stopping. This labeling model is used to generate pseudo-labels for unlabeled audio in Stage 2.

\paragraph{Stage 2: Continued Pretraining}
Using the labeling model, we generate pseudo-labels for unlabeled Swahili audio via greedy CTC decoding, retaining only samples with confidence scores above 75\%. This filtering ensures the pseudo-labels provide useful training signal while limiting noise. We then continue pretraining wav2vec2-bert-2.0 using these pseudo-labels as supervision targets.

For training, we adopt conservative hyperparameters to prevent catastrophic forgetting: learning rate 5e-5, only 3 epochs, batch size 8, warmup ratio 10\%, and AdamW optimizer~\cite{loshchilov2019adamw} with weight decay 0.01.

\paragraph{Stage 3: Supervised Finetuning}
We start from the CPT checkpoint and finetune on labeled Common Voice data (5K or 20K samples). We use learning rate 1e-4, batch size 8, and train for 10-15 epochs depending on configuration. We apply regularization techniques including label smoothing (0.1), dropout, and gradient clipping, with early stopping monitoring validation WER (patience 3).

\section{Results}

\subsection{Main Findings}

Table~\ref{tab:main_results} presents our complete results across all training stages and configurations.

\begin{table}[h]
\centering
\caption{Swahili ASR performance across training configurations. Baseline refers to the 50K comparison model trained without CPT. $\Delta$ represents relative improvement over baseline.}
\label{tab:main_results}
\small
\begin{tabular}{lcccc}
\toprule
Config & Baseline & Final WER & $\Delta$ & Benefit \\
\midrule
5K+CPT & 17.71\% & \textbf{10.89\%} & -38.5\% & \checkmark \\
20K+CPT & 17.71\% & \textbf{3.24\%} & -81.7\% & \checkmark \\
\bottomrule
\end{tabular}
\end{table}

Several key observations emerge from these results:

\begin{itemize}
\item Both configurations achieve substantial improvements over the comparison baseline, with 20K+CPT reaching 3.24\% WER
\item The 20K+CPT configuration substantially outperforms the 50K comparison baseline (3.24\% vs 17.71\%), demonstrating that CPT with less labeled data produces markedly better results than supervised training alone with more labeled data
\item The 3.24\% WER surpasses all previously published academic Swahili benchmarks
\item Appropriate training strategies matter more than massive labeled datasets---achieving state-of-the-art with just 20K samples
\end{itemize}

\subsection{Detailed Analysis}

\subsubsection{5K Configuration}

The comparison baseline (50K samples, no CPT) achieves 17.71\% WER. With the CPT approach using just 5K labeled samples, the model achieves 10.89\% WER---an improvement of 6.82 percentage points representing a 38.5\% relative reduction. With just 3 hours of labeled data, our CPT approach enables competitive performance suitable for various applications including draft transcription, voice search, and language learning tools.

\subsubsection{20K Configuration}

With the CPT approach using 20K labeled samples, the model achieves 3.24\% WER---representing 14.47 percentage points improvement or 81.7\% relative reduction from the comparison baseline. With 11 hours of labeled data, our CPT approach establishes new state-of-the-art for Swahili ASR, surpassing the best academic model (XLS-R's 8.3\%) by 61\% relative improvement.

\subsection{Benchmark Comparison}

Table~\ref{tab:benchmarks} positions our results within the published Swahili ASR landscape.

\begin{table}[h]
\centering
\caption{Swahili ASR benchmarks comparing our approach with published academic systems}
\label{tab:benchmarks}
\small
\begin{tabular}{llc}
\toprule
System & Type & WER \\
\midrule
XLS-R finetuned~\cite{babu2021xls} & Academic & 8.3\% \\
SpeechBrain~\cite{ravanelli2021speechbrain} & Academic & 9-12\% \\
\textbf{Ours (5K+CPT)} & \textbf{Research} & \textbf{10.89\%} \\
\textbf{Ours (20K+CPT)} & \textbf{Research} & \textbf{3.24\%} \\
\bottomrule
\end{tabular}
\end{table}

\textbf{Significance:}
\begin{itemize}
\item \textbf{New state-of-the-art}: Best reported WER on Common Voice Swahili among academic systems
\item \textbf{Surpasses academic baselines}: Exceeds XLS-R by 61\% relative improvement
\item \textbf{Accessible methodology}: Uses publicly available models and datasets
\item \textbf{Demonstrates feasibility}: Shows that high-quality Swahili ASR is achievable with modest resources
\end{itemize}

\section{Discussion}

Our results demonstrate that continued pretraining on pseudo-labeled unlabeled data enables state-of-the-art Swahili ASR with minimal labeled data requirements. The 3.24\% WER achieved with just 20,000 labeled samples represents a 61\% improvement over the best academic system (XLS-R's 8.3\%).

\subsection{Why Does CPT Work?}

Several factors contribute to CPT's success:

\subsubsection{Strong Labeling Model Enables Quality Pseudo-Labels}

A labeling model with WER below 25\% provides sufficiently accurate pseudo-labels to yield useful training signal without overwhelming the model with noise. This contrasts with scenarios where labeling model WER $>$40\%, where pseudo-label noise may dominate.

\subsubsection{Domain Diversity from Unlabeled Audio}

The unlabeled audio provides exposure to diverse conditions: diverse speakers (age, gender, regional variation); multiple domains (news, education, entertainment); varied acoustic conditions (studio, outdoor, mobile); natural speech patterns (vs. read speech in Common Voice).

\subsubsection{Objective Alignment}

Both CPT (pseudo-labeled) and supervised finetuning use CTC loss with character-level predictions. This objective alignment facilitates smooth transition between stages.

\subsubsection{Conservative Hyperparameters}

Our conservative approach (learning rate 5e-5, only 3 epochs) limits how much CPT can disrupt pretrained representations, successfully balancing adaptation against forgetting.

\subsection{Practical Guidelines}

Based on our findings, we provide recommendations for practitioners:

\textbf{Requirements for CPT:}
\begin{itemize}
\item At least 20K labeled samples available to train an initial baseline model
\item Baseline model achieves WER $<$25\%, sufficient for generating quality pseudo-labels
\item Unlabeled target language audio is readily available
\end{itemize}

\textbf{Data Collection Strategy:}

Phase 1: Train a baseline model on labeled data ($\geq$20K samples) to achieve WER $<$25\%.

Phase 2: Collect unlabeled audio and generate pseudo-labels using the baseline model, filtering for high-confidence outputs.

Phase 3: Continue pretraining on the pseudo-labeled data, then finetune on labeled data.

Total Investment: The labeled data requirement is modest compared to the hundreds of hours typically required for traditional supervised ASR systems.

\section{Conclusion}

We have demonstrated that continued pretraining on pseudo-labeled unlabeled data enables state-of-the-art Swahili ASR with minimal labeled data requirements. Our key contributions include:

\begin{enumerate}
\item \textbf{First systematic evaluation} of pseudo-labeled CPT for Swahili
\item \textbf{New state-of-the-art} for Swahili ASR (3.24\% WER)
\item \textbf{Concrete data requirements}: $\sim$20K labeled samples combined with unlabeled audio for high-quality deployment
\item \textbf{Practical, replicable methodology} applicable to other low-resource languages
\end{enumerate}

Our 3.24\% WER represents a 61\% improvement over the best academic system. This is the best reported WER on Common Voice Swahili among academic systems, achieved with just 20K labeled samples.

The broader impact extends to over 100 million Swahili speakers who can benefit from high-quality ASR technology, enabling educational technology in mother tongue, accessibility tools, voice interfaces, and documentation of oral traditions. Our results demonstrate that high-quality ASR is achievable for low-resource languages given appropriate training methodology, modest labeled data resources, and readily-available unlabeled audio.

\balance  

\bibliographystyle{IEEEtran}
\bibliography{references}

\end{document}